\documentstyle[aps,prl,twocolumn,epsf]{revtex}
\begin{document}
\input{psfig.sty}
\draft
\twocolumn[\hsize\textwidth\columnwidth\hsize\csname 
@twocolumnfalse\endcsname 

\title{Soft Mode Dynamics Above and Below the Burns Temperature in the
  Relaxor Pb(Mg$_{1/3}$Nb$_{2/3}$)O$_3$}

\author{P. M. Gehring,$^1$ S. Wakimoto,$^{2,3}$ Z.-G. Ye,$^4$ and G.
  Shirane$^3$}

\address{$^1$NIST Center for Neutron Research, National Institute of
  Standards and Technology, Gaithersburg, Maryland 20899}

\address{$^2$Department of Physics, University of Toronto, Toronto,
  Ontario, Canada M5S 1A7}

\address{$^3$Physics Department, Brookhaven National Laboratory,
  Upton, New York 11973}

\address{$^4$Department of Chemistry, Simon Fraser University, Burnaby,
  British Columbia, Canada V5A 1S6}

\date{\today}
\maketitle

\begin{abstract}
  We report neutron inelastic scattering measurements of the
  lowest-energy transverse optic (TO) phonon branch in the relaxor
  Pb(Mg$_{1/3}$Nb$_{2/3}$)O$_3$ from 400 to 1100~K.  Far above the
  Burns temperature $T_d \sim 620$~K we observe well-defined
  propagating TO modes at all wave vectors $q$, and a zone center TO
  mode that softens in a manner consistent with that of a
  ferroelectric soft mode.  Below $T_d$ the zone center TO mode is
  overdamped.  This damping extends up to, but not above, the
  waterfall wave vector $q_{wf}$, which is a measure of the average
  size of the PNR.
\end{abstract}


]

The lead-oxide class of relaxor compounds has received an enormous
amount of attention from the scientific and industrial communities in
the past several years due to their extraordinary piezoelectric
properties and the corresponding promise they hold for device
applications.  Of particular interest are the prototypical relaxor
Pb(Mg$_{1/3}$Nb$_{2/3}$)O$_3$ (PMN), and the closely related system
Pb(Zn$_{1/3}$Nb$_{2/3}$)O$_3$ (PZN), both of which show a huge
increase in piezoelectric character when doped with the normal
ferroelectric perovskite PbTiO$_3$ (PT) \cite{Park}.  Despite intense
research efforts, no consensus on a fundamental understanding of the
basic lattice dynamics of these relaxor systems has been established.
The identification of a soft transverse optic (TO) mode at the
Brillouin zone center, for example, the hallmark of every displacive
ferroelectric phase transition including that which occurs in
PbTiO$_3$ at 763~K \cite{Gen}, has proven elusive in both PMN and PZN.

In an effort to solve the soft mode puzzle in the relaxor compounds,
we have performed an extensive study of the lattice dynamics of pure
PMN using neutron inelastic scattering techniques at temperatures far
above $T_{max} = 265$~K, where the dielectric susceptibility exhibits
a broad peak.  The temperature scale for this study is set by the
Burns temperature $T_d \sim 620$~K for PMN, where randomly-oriented
regions of local polarization, of order several unit cells in size,
begin to condense within the otherwise non-polar crystal structure.
Experimental evidence for these so-called polar nanoregions (PNR) was
first obtained by Burns and Dacol in 1983 from measurements of the
optic index of refraction on both PMN and PZN, as well as other
disordered systems \cite{Burns}.  Previous neutron inelastic studies
on single crystals of PZN and PZN doped with 8\% PbTiO$_3$ (PZN-8\%PT)
in their respective cubic phases at 500~K have shown that the
lowest-energy TO phonon modes are overdamped for reduced wave vectors
$q$ less than a characteristic wave vector $q_{wf}$, which is of order
0.2~\AA$^{-1}$, but underdamped otherwise \cite{Gehring1,Gehring2}.
It was speculated that the PNR are the underlying cause of this
$q$-dependent damping because the TO modes are polar, and thus they
should couple strongly to the PNR.  Indeed, it is this coupling that
is believed to produce the anomalous waterfall feature observed in
these and other lead-oxide relaxor systems including PMN
\cite{Gehring3,ASR}.  If these speculations are correct, then the
search for a true soft mode can only be made above $T_d$ since the
anomalous low-$q$ damping makes it impossible to observe the
temperature dependence of the zone center TO mode.  However, the
values of $T_d$ for PZN and PZN-$x$PT compounds approach their
decomposition temperatures, making such measurements difficult.  PMN,
by contrast, has a much lower Burns temperature compared to that of
PZN, making it an ideal system in which to look for the soft mode.  We
have exploited this fact to study the evolution of the lowest-energy
TO branch in PMN at temperatures far above and below $T_d$, thereby
providing key insight into the role played by the PNR on the soft mode
dynamics in this important relaxor system.

The inset to Fig.~1 shows the TO and TA dispersion curves measured
along the cubic [001] direction after heating the PMN crystal in
vacuum to 1100~K, well above $T_d$, but safely below the decomposition
temperature of $\sim$ 1370~K.  The data shown in panels (a) and (b)
are constant-$\vec{Q}$ scans measured at $\vec{Q}$ = (2,0,0.08) and
(2,0,0.16), respectively, and indicate that well-defined propagating
TO and TA modes are present throughout the Brillouin zone at this
temperature.  Similar data measured at 800~K were discussed briefly by
Naberezhnov {\it et al.} who identified the upper curve as a hard TO1
branch, and concluded that it could not be the ferroelectric mode
\cite{Naberezhnov}.  Based on our PMN data, which cover a larger range
in temperature, this conclusion does not appear to be justified.
Instead, we demonstrate in this Letter that the lowest-frequency TO
mode is in fact the elusive ferroelectric soft mode in this relaxor
compound, and that the PNR have a drastic effect on its behavior.

%
%
\begin{figure}
\begin{center}
\parbox[b]{3.375in}{\psfig{file=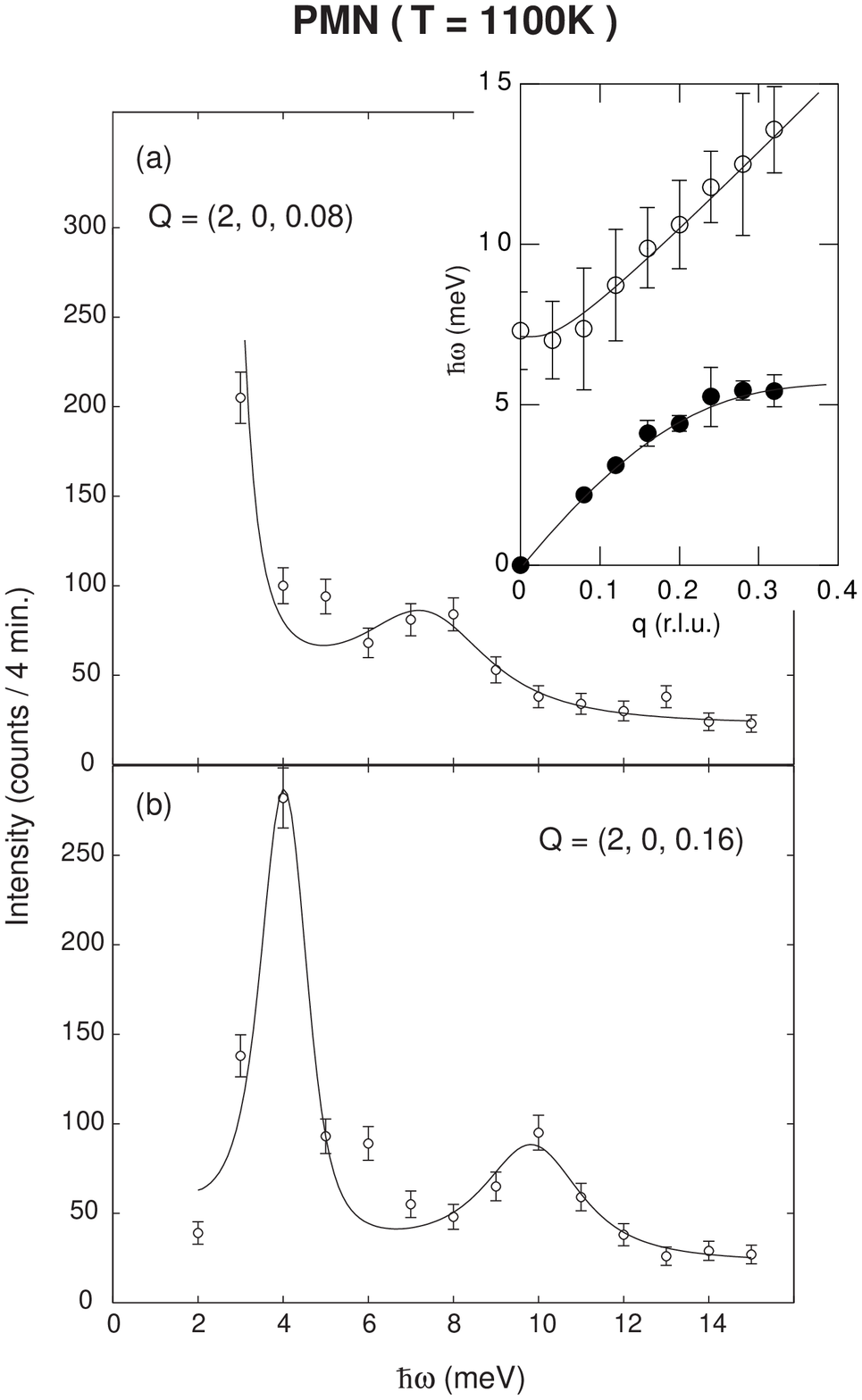,width=2.75in}
  {\vspace{0.1in} Fig.\ 1.  \small Constant-$\vec{Q}$ scans measured
    at 1100~K ($>T_d$).  Panel (a) shows a well-defined, underdamped
    TO mode for $q = 0.08$~rlu.  Panel (b) shows well-defined,
    underdamped TO and TA modes for $q = 0.16$~rlu.  The TA and TO
    dispersion curves are shown in the inset (lines are guides to the
    eye). }}
\end{center}
\label{fig:1}
\end{figure}
%
%

The neutron inelastic scattering data presented here were obtained on
the BT9 triple-axis spectrometer located at the NIST Center for
Neutron Research.  We used the same experimental configuration as that
described in Ref.~[4].  The data were taken holding the final neutron
energy $E_f$ fixed at 14.7~meV ($\lambda_i = 2.36$~\AA) while varying
the incident neutron energy $E_i$.  Horizontal beam collimations were
40$'$-46$'$-S-40$'$-80$'$.  Constant-$E$ scans were performed by
holding the energy transfer $\hbar \omega = E_i - E_f$ fixed while
varying the momentum transfer $\vec{Q}$.  Constant-$\vec{Q}$ scans
were performed by holding the momentum transfer $\vec{Q} = \vec{k_i} -
\vec{k_f}$ ($k = 2\pi/\lambda$) fixed while varying $\hbar\omega$.

Single crystals of PMN were grown from high temperature solution using
PbO as flux.  The growth conditions were determined from the
pseudo-binary phase diagram established for PMN and PbO \cite{Ye}.  A
rectangular parallelepiped crystal, with dimensions $8.7 \times 5.1
\times 2.2$ mm$^3$ (0.10 cm$^3$), was prepared with the largest facets
oriented parallel to the cubic [100] direction.  This crystal was
mounted in a molybdenum holder with quartz wool with the [010] axis
oriented vertically, and attached to the end of a long sample stick.
The sample stick assembly was then mounted inside the vacuum space of
a furnace, and positioned onto the goniometer of the spectrometer.
This orientation gave access to reflections of the form $(h0l)$.

%
%
\begin{figure}
\begin{center}
\parbox[b]{3.375in}{ \psfig{file=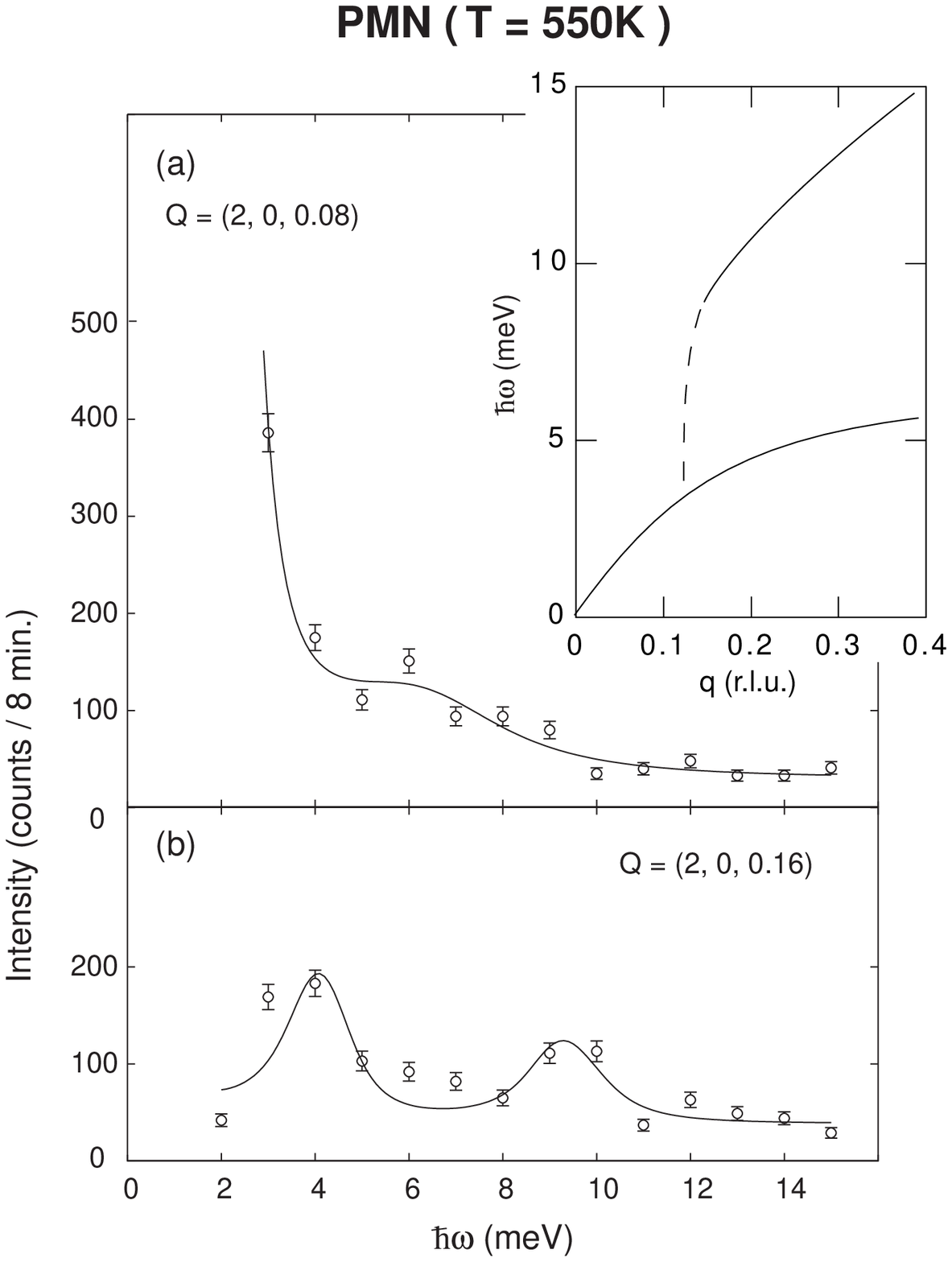,width=2.75in}
  { \vspace{0.1in} Fig.\ 2.  \small Constant-$\vec{Q}$ scans measured
    at 550~K ($< T_d$).  Panel (a) shows a damped TO mode for $q =
    0.08$~rlu $< q_{wf}$.  Panel (b) shows well-defined, underdamped
    TA and TO modes for $q = 0.16$~rlu $> q_{wf}$.  The inset shows
    the TA and TO dispersion curves below $T_d$, the latter of which
    terminates at the waterfall wave vector (dashed line) at $q_{wf}
    \sim$ 0.12~rlu.}}
\end{center}
\label{fig:2}
\end{figure}
%
%

The data presented in Fig.~2 were taken below $T_d$, but still in the
cubic phase of PMN, to establish consistency with prior results on PZN
and PZN-8\%PT \cite{Gehring1,Gehring2}.  The solid curves shown in the
inset were derived from constant-$\vec{Q}$ scans, whereas the dashed
line representing the waterfall anomaly was derived from constant-$E$
scans.  The waterfall feature is located at $q = q_{wf} \sim 0.12$~rlu
(reciprocal lattice units), where 1 rlu = $2\pi/a$ = 1.553~\AA$^{-1}$.
Fig.~2 (a) shows a representative constant-$\vec{Q}$ scan measured at
$\vec{Q} = (2,0,0.08)$ ($q < q_{wf}$) where, in contrast to that shown
in Fig.~1 (a), the TO mode is now strongly damped and appears at lower
energy.  The solid lines in these and all other panels represent fits
of the data to a Lorentzian function of $q$ and $\omega$ convoluted
with the instrumental resolution function.  The TA mode is not visible
in panel (a) because it peaks at energies below 3~meV.  Fig.~2 (b)
shows a similar constant-$\vec{Q}$ scan measured at $\vec{Q} =
(2,0,0.16)$ ($q > q_{wf}$) for which well-defined propagating TA and
TO modes are observed.  The peak intensities, which scale with the
Bose factor, are noticeably lower for the TA mode in Fig.~2 (b)
compared to the TA mode in Fig.~1 (b).  (To facilitate comparison
between figures, all intensity axes have been scaled to the same count
rate per unit length.)

A series of constant-$\vec{Q}$ scans were taken in steps of 0.04~rlu
over the entire Brillouin zone, and the results are summarized in the
inset to Fig.~2.  They show that at 550~K propagating TO modes are now
only present for wave vectors $q > q_{wf} \sim$ 0.12~rlu, and that the
waterfall feature has appeared.  The dashed line in the inset is used
to indicate that the waterfall is in fact {\it not} part of the TO
dispersion curve since no propagating modes are observed for $q \le
q_{wf}$.  We emphasize that these data were taken at 550~K, far above
$T_{max} = 265$~K, the temperature at which the dielectric
susceptibility of PMN reaches a maximum, yet still below the Burns
temperature $T_d \sim 620$~K.  Hence while the system is in the cubic
phase, it also contains a finite density of nanometer-sized regions of
randomly-oriented polarization.  The data taken on PMN at this
temperature are thus consistent with the picture obtained from earlier
neutron studies of PZN and PZN-8\%PT.  The question remains, however,
of whether or not this damping, and the resulting waterfall, correlate
with the condensation of the PNR at the Burns temperature.

To answer this question, we show three identical constant-$\vec{Q}$
scans in Fig.~3 measured at the zone center at $\vec{Q} = (2,0,0)$
from 1100~K (above $T_d$) down to 600~K (near $T_d$).  These data
reveal an unambiguous evolution from an underdamped to an overdamped
phonon cross section as the temperature is reduced below $T_d$.  The
TO peak intensity diminishes, and the TO linewidth broadens, with
decreasing temperature.  Although the damping of the zone center mode
begins at temperatures considerably above $T_d$, we note that the mode
becomes completely overdamped near or at $T_d$ as shown in the bottom
panel of Fig.~3.  The increase in the damping of this and other
low-$q$ modes is accompanied by the gradual formation of the
waterfall.  It appears that $T_d$ coincides with the condensation of
the dynamical fluctuations of the PNR at $q=0$ which develop at much
higher temperatures.

The temperature dependence of the zone center TO mode energy squared
$(\hbar\omega_0)^2$, shown in the inset to Fig.~3, varies linearly
with temperature until $T \sim T_d$.  This behavior is consistent with
that observed in PbTiO$_3$ \cite{Gen} and other displacive
ferroelectrics, and represents the first definitive identification of
a true soft mode in PMN.  At present we have no direct experimental
measure of the zone center mode frequency below $T_d$.  However, we
speculate that this mode recovers at very low temperature as has been
reported for PZN where, at 20~K, the TO mode scattering cross section
peaks at 10.5~meV \cite{Gehring2}.  A dashed line connects this data
point to that for PMN at $T_d$ to suggest how this mode might recover
at low temperature.

To establish the temperature dependence of the TO mode damping in
greater detail, we present data in Fig.~4 similar to those shown in
Fig.~3, but for non-zero $q$.  As was the case for the zone center
mode, the TO mode at $q = 0.08$~rlu gradually softens with decreasing
temperature above $T_d$.  Moreover, the top and bottom panels of
Fig.~4 indicate that the TO mode is underdamped at high temperature
and overdamped at low temperature.  However, the presence of a weak
phonon peak at 550~K (below $T_d$) in the middle panel of Fig.~4
demonstrates the important fact that TO modes with finite $q$ become
overdamped at lower temperatures compared to that for the zone center
mode.  Hence the temperature dependence of the TO mode damping is
$q$-dependent.

%
%
\begin{figure}
\begin{center}
\parbox[b]{3.375in}{\psfig{file=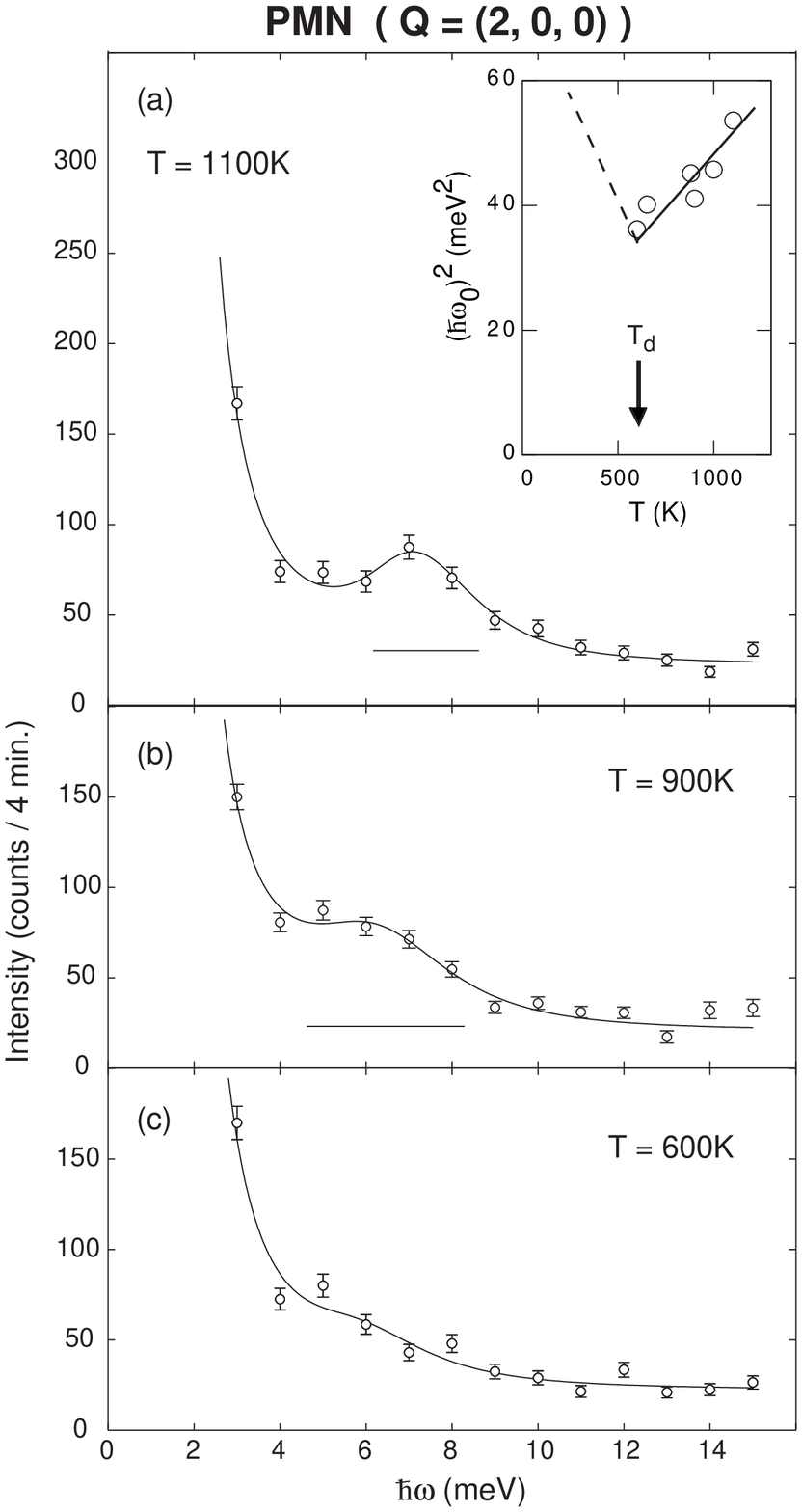,width=2.75in}
  {\vspace{0.1in} Fig.\ 3.  \small The square of the zone center TO
    mode energy $(\hbar\omega_0)^2$ softens linearly (see inset)
    between (a) 1100~K, (b) 900~K, and (c) 600~K.  The FWHM linewidth
    (horizontal bar) increases from 1.2~meV at 1100~K to 1.8~meV at
    900~K, but is too broad to measure reliably at 600~K.}}
\end{center}
\label{fig:3}
\end{figure}
%
%

We have identified a soft polar phonon mode in PMN, and established
its behavior through the Burns temperature $T_d$.  Our identification
of the lowest-energy zone center TO mode as the ferroelectric soft
mode differs from that of Naberezhnov {\it et al}.\ who identified it
as a hard TO1 mode \cite{Naberezhnov}.  In their neutron study, they
claimed that no underdamped soft TO mode was observed in the vicinity
of (221), even at 900~K and large $q$.  However the data we present in
Fig.'s 3 -- 4 show conclusively that our mode determination is
correct.  Indeed, the relative intensities of the TO branch measured
in different zones are very similar to those found in PbTiO$_3$.  It
is important to note that the existence and value for $T_d$ has been
confirmed by a variety of experimental techniques.  Neutron powder
diffraction data by Zhao {\it et al.}, for example, provide clear
evidence of the PNR in PMN through a marked deviation from the linear
dependence on temperature of the unit cell volume near 600~K
\cite{Zhao}.  Also, Naberezhnov {\it et al}.\ observe diffuse
scattering at (2,2,0.95) that begins to increase at 650~K, as well as
a TA linewidth broadening at the same temperature \cite{Naberezhnov}.
Our measurements indicate a Burns temperature that lies between 600
and 650~K, and are thus consistent with both of these studies.

%
%
\begin{figure}
\begin{center}
\parbox[b]{3.375in}{\psfig{file=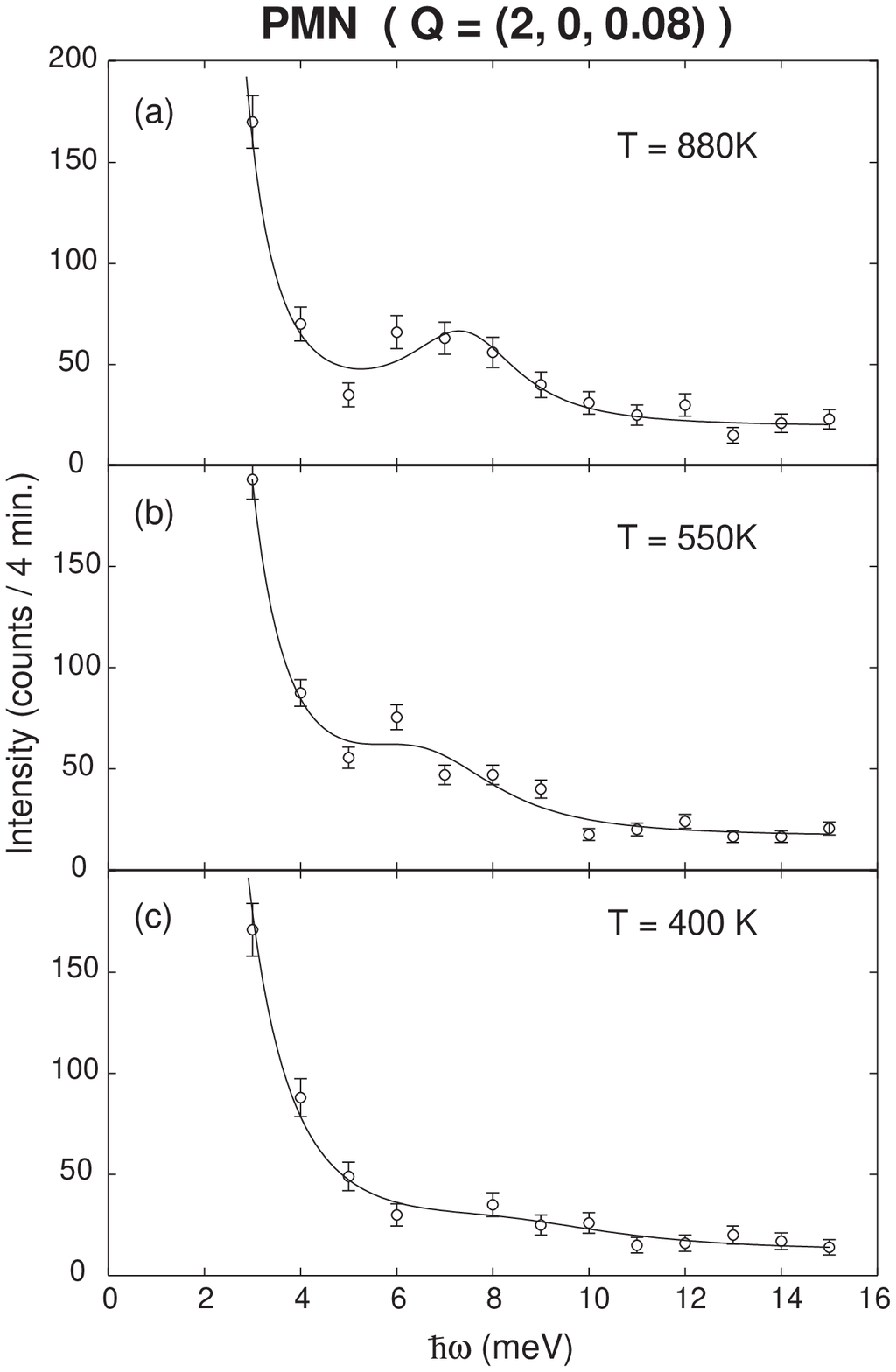,width=2.75in}
  {\vspace{0.1in} Fig.\ 4.  \small The TO mode at $\vec{Q} =
    (2,0,0.08)$ is shown for (a) 880~K, (b) 550~K, and (c) 400~K.
    This mode also becomes overdamped (bottom panel), but only {\em
      below} $T_d \sim 620$~K. }}
\end{center}
\label{fig:4}
\end{figure}
%
%

Our next goal is to determine how the diffuse scattering in PMN
connects to this soft mode picture.  Diffuse scattering was first
reported in PMN by Vakhrushev {\it et al.} \cite{Vakhrushev-N} at room
temperature, however the scattering intensities are entirely different
from those of the soft optic mode.  For example, the scattering near
(110) is much stronger than that at (200).  In fact, this observation
was the main reason that Naberezhnov {\it et al}.\ assigned the
lowest-frequency TO branch as a hard TO1 branch \cite{Naberezhnov,QO}.
Although several x-ray studies of the diffuse scattering in PMN have
been published \cite{Vakhrushev-X,You,Takesue}, the ionic shifts
derived from the neutron diffuse scattering peaks in PMN
\cite{Vakhrushev-N} have not yet been confirmed by x-ray scattering.
We are trying to reconcile the conflict between the diffuse and soft
mode scattering cross sections by considering different models.
Finally, the question of how the long-wavelength TO phonon modes
recover at low temperatures, as was observed in PZN, remains to be
answered.  Neutron experiments using a larger PMN crystal are planned
to examine this interesting issue.

We thank Y.\ Fujii, K.\ Hirota, V.\ Kiryukhin, T.\ -Y.\ Koo, K.\ 
Ohwada, N.\ Takesue, and H.\ You for stimulating discussions.  We also
thank S.\ M.\ Shapiro and S.\ B.\ Vakrushev for sharing their PMN data
with us prior to publication.  We acknowledge financial support from
the U.\ S.\ Dept.\ of Energy under contract No.\ DE-AC02-98CH10886, as
well as the Office of Naval Research, Grant No.\ N00014-99-1-0738.
Work at the University of Toronto is part of the Canadian Institute
for Advanced Research and is supported by the Natural Science and
Engineering Research Council of Canada.


\begin{references}
\bibitem{QO} A fictitious ``quasi-optic'' branch was introduced,
  derived solely from the overdamped modes, in the context of their
  mode-coupling analysis (see Ref.\ [6]).
\end{references}
\end{document}